\documentclass[a4paper,10pt,twoside]{cpc-hepnp}

\usepackage{multicol}
\usepackage{graphicx}
\usepackage{booktabs}
\usepackage{amssymb,bm,mathrsfs,bbm,amscd}
\usepackage[tbtags]{amsmath}
\usepackage{lastpage}

\begin{document}

\fancyhead[c]{\small Chinese Physics C~~~Vol. xx, No. x (201x) xxxxxx}
\fancyfoot[C]{\small 010201-\thepage}


\title{Screening Effects on the Binding Energy and Stability of Quarkonia States}

\author{%
      Palak Bhatt$^{1}$,\email{palak.physics@gmail.com}%
\quad Arpit Parmar$^{2}$,\email{arpitspu@yahoo.co.in}%
\quad Smruti Patel$^{1}$,\email{fizixsmriti31@gmail.com}%
\quad P. C. Vinodkumar$^{1}$\email{pc$\_$vinodkumar@spuvvn.edu}%
}
\maketitle

\address{%
$^1$ Department of Physics, Sardar Patel University,Vallabh Vidyanagar-388120, INDIA.\\
$^2$ {Nuclear Physics Devision, Bhabha Atomic Research Centre, Mumbai-400085, India. } \\
}

\begin{abstract}
We have studied the thermal stability of Quarkonia states by computing the effects of color-screening and vacuum screening based on a temperature dependent screened coulomb plus power potential for the quark-antiquark interaction. Medium effects on the properties of charmonia and bottomonia states are studied. The color screening and the vacuum screening effects on the stability of the quarkonia states are also separately calculated for comparison.
\end{abstract}



\footnotetext[0]{\hspace*{-3mm}\raisebox{0.3ex}{$\scriptstyle\copyright$}2013
Chinese Physical Society and the Institute of High Energy Physics
of the Chinese Academy of Sciences and the Institute
of Modern Physics of the Chinese Academy of Sciences and IOP Publishing Ltd}%

\begin{multicols}{2}

\section{Introduction} \label{intro}

A fundamental description of the behavior of particles in a thermal environment can be represented through screening mass as it considers the interaction of a particle with the medium. Such an understanding about the screening mass of the elementary particles (quark, gluon, leptons, $W^{\pm}$-bosons, Z-bosons, higgs bosons) is necessary for understanding the formation of deconfined quark-gluon plasma state (the matter with high density at high temperature) which has been intensely studied in heavy ion collision experiments at CERN, BNL and RHIC etc. It also plays an important role at high temperature in strong and electroweak interaction and also in understanding the deconfinement mechanism. The parton jets, electromagnetic signals and the suppression of quarkonia states are the main signatures of the deconfined state\cite{NA50,NA60,PHENIX}. Though there exist many theory like the effective field theory and lattice simulations to study the deconfinement mechanisms but they require very intense computational efforts. However study based on phenomelogical model is very simple and an important tool to understand the screening effects on the binding energy of quarkonia states.

\section{formalism}

Study of the deconfined medium has been attempted using the schrodinger equation with a non-relativistic Hamiltonian given by
\begin{equation}\label{H}
H=M+\frac{p^{2}}{2m}+ V(r,T)
\end{equation}
where, M=$m_{1}$+$m_{2}$ and $m$=$\frac{m_{1} m_{2}}{m_{1}+m_{2}}$. Here, $m_{1/2}$ corresponds to the mass of the quark/antiquark constituting the quarkonia state. For example, in the case of charmonium, $m_{c}$=1.320GeV/$c^{2}$ and in the case of bottomonium $m_{b}$=4.746GeV/$c^{2}$ \cite{Karsch,Satz_}. The medium dependent quark-antiquark potential\cite{arpit} is considered as
\begin{equation}\label{V(mu)}
V(r,\mu(T))=\frac{-\alpha}{r} exp[-\mu(T)r]+ \frac{\sigma}{\mu(T)}(1-exp[-\mu(T)r^{\nu}])
\end{equation}
where, $\alpha$=0.471\cite{Jacobs} and $\sigma$ has been determined by taking the corresponding spin average mass of charmonia and bottomonia (1s, 2s and 1p-states) without considering the medium effects ($\mu\rightarrow$0). The parameter $\sigma$ for different choices of $\nu$ thus are obtained are plotted in the fig.(\ref{sigma_c}) and (\ref{sigma_b}) for the charmonia and bottomonia respectively. It is found that for $\nu$=1 the value of $\sigma$ remains almost same for both $c\bar{c}$ and $b\bar{b}$ states.

\includegraphics[width=6cm, height=5cm]{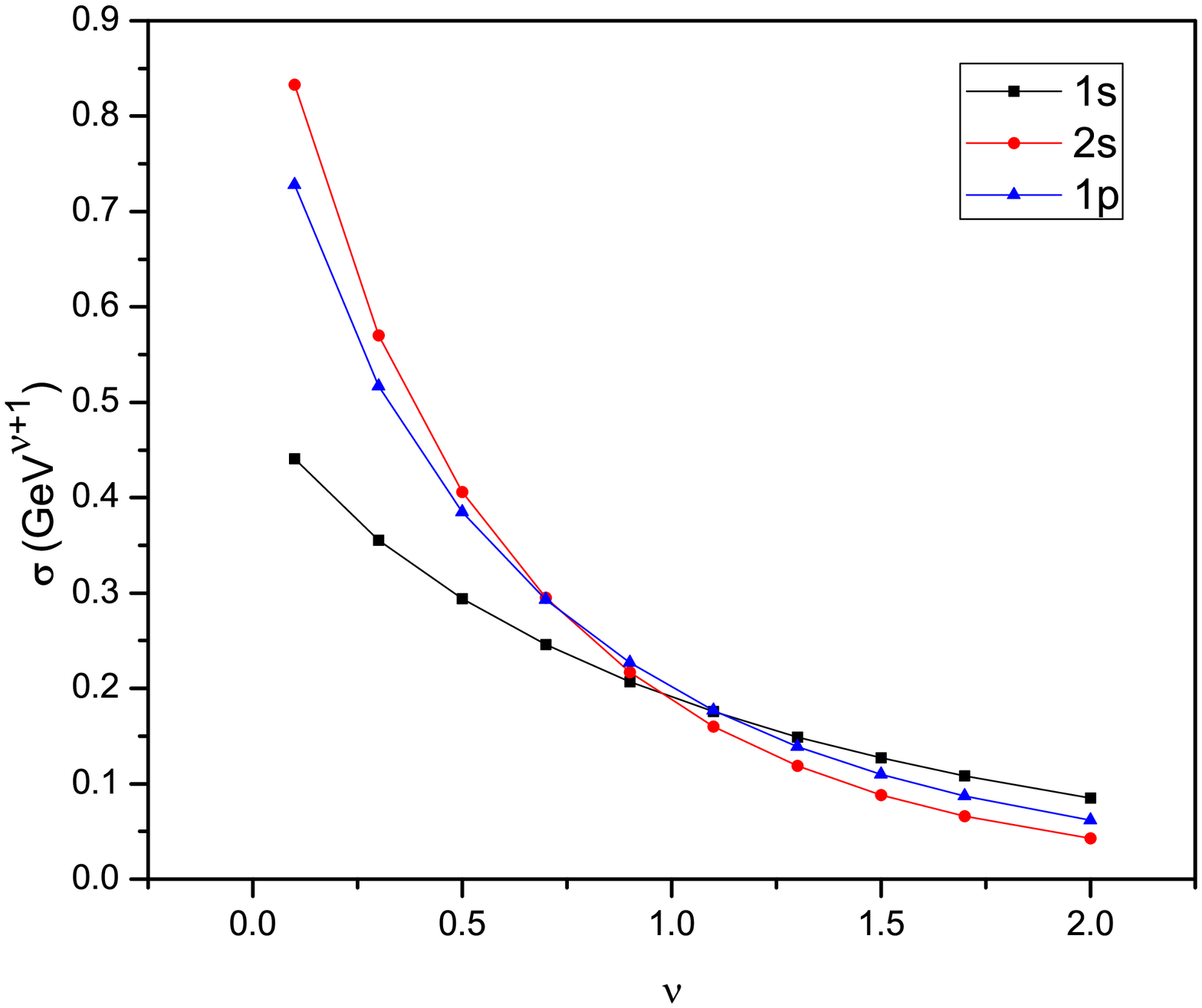}
\figcaption{string tension $\sigma$ at different choices of power index $\nu$ for charmonia}\label{sigma_c}

\includegraphics[width=6cm, height=5cm]{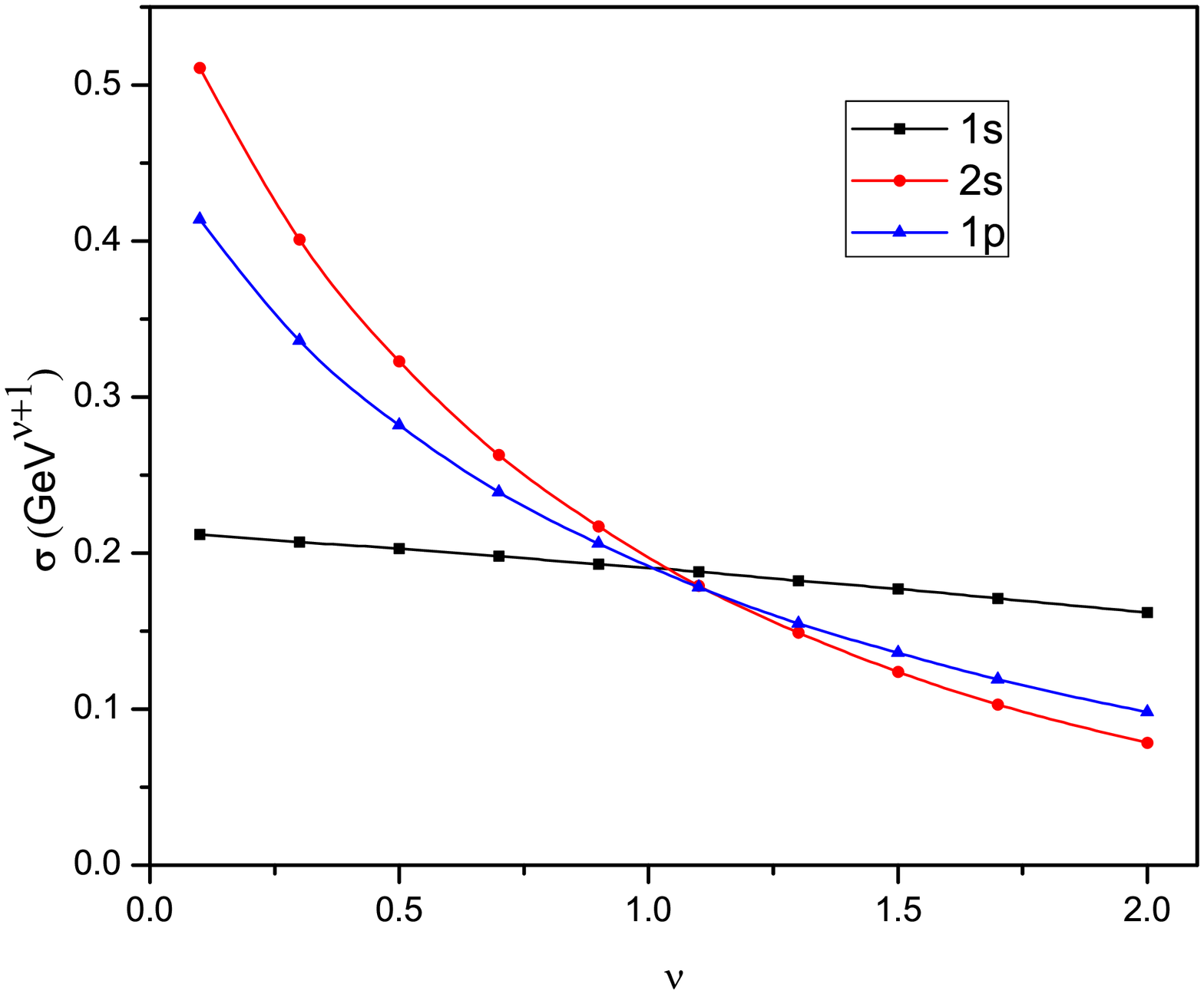}
\figcaption{string tension $\sigma$ at different choices of power index $\nu$ for bottomonia}\label{sigma_b}
\end{multicols}

\includegraphics[width=6cm, height=6cm]{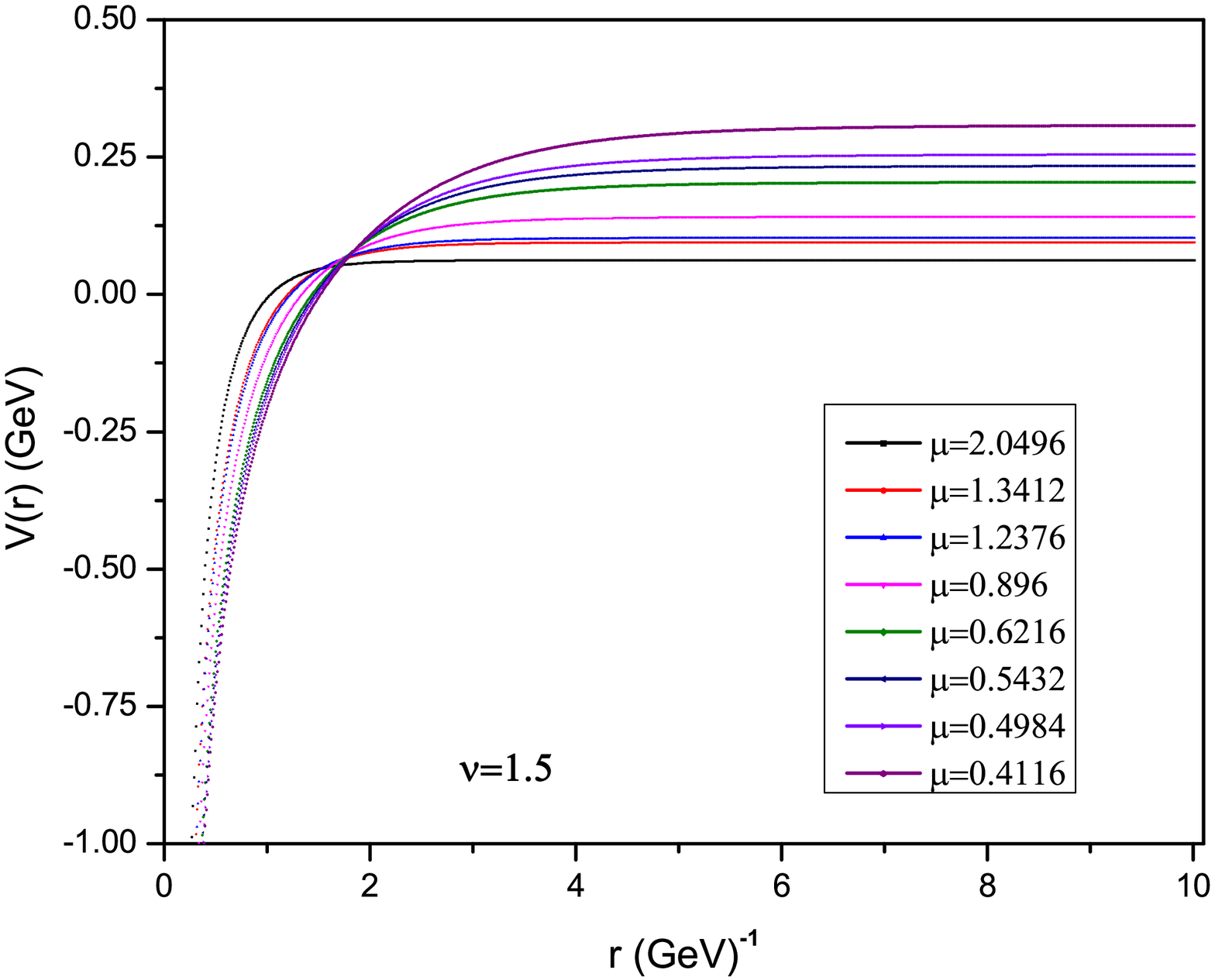}
\includegraphics[width=6cm, height=6cm]{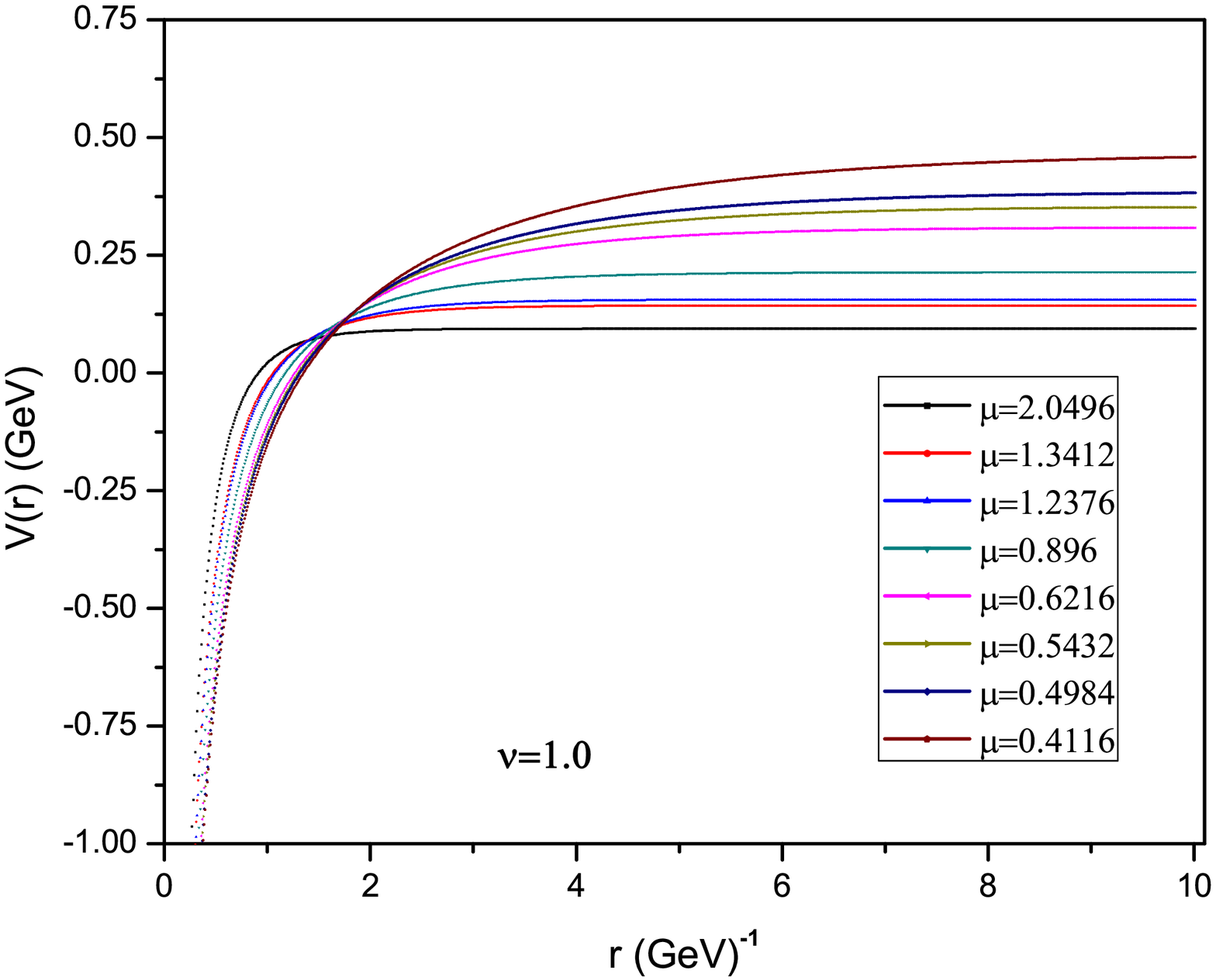}
\includegraphics[width=6cm, height=6cm]{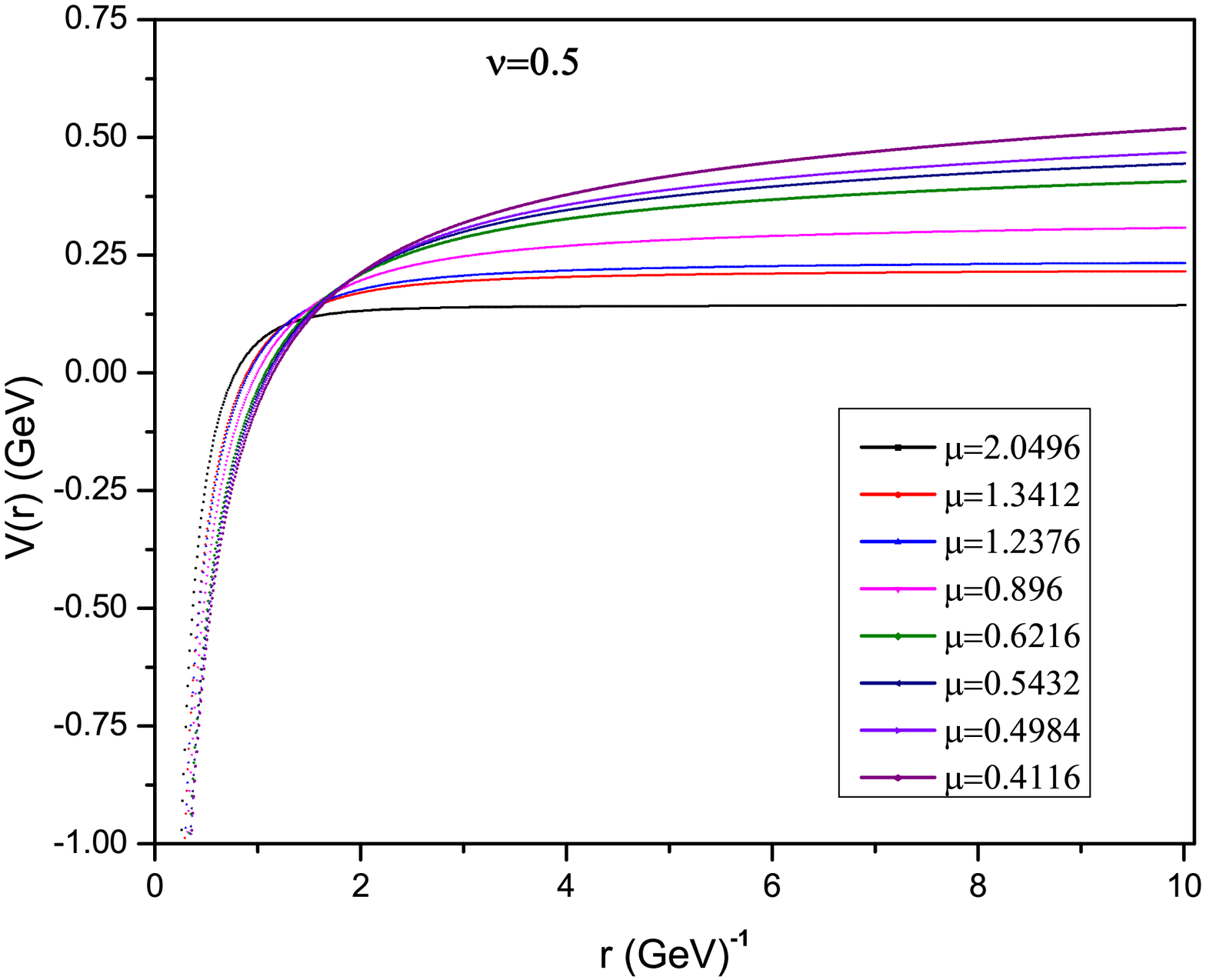}
\figcaption{The potential model from eqn.(\ref{V(mu)}) for particular choices of $\mu$ and $\nu$}\label{vr_r}

\begin{multicols}{2}

The nature of potential $\ref{V(mu)}$ for the different choices of $\nu$=0.5, 1.0 and 1.5 is showed in the fig.(\ref{vr_r}).

\section{Properties of Quarkonia states with medium effects $\mu\neq$0}

The Schr$\ddot{o}$dinger equation with the potential defined by eqn. (\ref{V(mu)}),
\begin{equation}\label{schr_eqn}
[\frac{1}{2m}(\frac{-d^{2}}{dr^{2}}+\frac{l(l+1)}{r^{2}})+V(r,\mu)]\Phi_{n,l}(r)= E_{n,l}(r) \Phi_{n,l}(r),
\end{equation}
is solved \cite{Lucha} to get the energy eigenvalue $E^{n,l}(\mu)$ as a function of the medium parameter, $\mu$. We now define an effective binding energy express as \cite{Karsch,Peter}
\begin{equation}\label{e_cs}
E^{n,l}_{cs}(\mu)\equiv2m + \frac{\sigma}{\mu} -E_{n,l}(\mu)
\end{equation}

$E^{n,l}_{cs}$($\mu$) described by eqn.(\ref{e_cs}) provides a positive value for the bound states and as $\mu$ increases, it decreases. For a particular value of $\mu=\mu_{c}$ at which

\begin{equation}\label{critical_mu}
E^{n,l}_{cs}(\mu=\mu_{c})=0
\end{equation}
defines a critical value for the screening mass $\mu_{c}$, beyond which no more binding is possible and it just dissociates.

\begin{center}
\tabcaption{Screening parameters of the $c\bar{c}$ (1s)state for $\mu$=$\mu_{c}$ for the different choices of $\nu$}\label{tt1}
\begin{tabular}{|cccccc|}
\hline
$\nu$	&	$\sigma$	&	$\mu_{c}$	&	$r_{0}$	&	M	&	$r_{D}$	\\
\hline
\hline
	&	$GeV^{\nu +1}$	&	GeV	&	fm	&	GeV	&	fm	\\
\hline											
0.1	&	0.441	&	1.01547	&	0.67704	&	3.07428	&	0.19433	\\
0.3	&	0.355	&	0.90442	&	0.72658	&	3.03252	&	0.21819	\\
0.5	&	0.294	&	0.80337	&	0.77376	&	3.00595	&	0.24564	\\
0.7	&	0.246	&	0.71047	&	0.81914	&	2.98624	&	0.27775	\\
0.9	&	0.207	&	0.62863	&	0.86436	&	2.96929	&	0.31392	\\
1.0	&	0.192	&	0.59417	&	0.88668	&	2.96314	&	0.33212	\\
1.1	&	0.176	&	0.5612	&	0.9136	&	2.95361	&	0.35163	\\
1.3	&	0.149	&	0.50597	&	0.97351	&	2.93448	&	0.39002	\\
1.5	&	0.127	&	0.46265	&	1.04001	&	2.9145	&	0.42654	\\
1.7	&	0.108	&	0.42787	&	1.1094	&	2.89241	&	0.46121	\\
2.0	&	0.085	&	0.3889	&	1.20957	&	2.85856	&	0.50743	\\
\hline
\end{tabular}
\end{center}

\begin{center}
\tabcaption{Screening parameters of the $c\bar{c}$ (2s)state for $\mu$=$\mu_{c}$ for the different choices of $\nu$}\label{tt2}
\begin{tabular}{|cccccc|}
\hline
$\nu$	&	$\sigma$	&	$\mu_{c}$	&	$r_{0}$	&	M	&	$r_{D}$	\\
\hline
\hline
	&	$GeV^{\nu +1}$	&	GeV	&	fm	&	GeV	&	fm	\\
\hline											
0.1	&	0.833	&	0.83193	&	0.96962	&	3.64128	&	0.23690	\\
0.3	&	0.57	&	0.70145	&	1.14599	&	3.45259	&	0.34620	\\
0.5	&	0.406	&	0.58339	&	1.35904	&	3.33593	&	0.48605	\\
0.7	&	0.295	&	0.47153	&	1.60487	&	3.26562	&	0.66894	\\
0.9	&	0.217	&	0.36723	&	1.86448	&	3.23091	&	0.90939	\\
1.0	&	0.192	&	0.32305	&	1.98204	&	3.23433	&	0.61086	\\
1.1	&	0.16	&	0.27809	&	2.14255	&	3.21535	&	1.23337	\\
1.3	&	0.119	&	0.21129	&	2.50665	&	3.2032	&	1.65831	\\
1.5	&	0.088	&	0.16417	&	3.04708	&	3.17603	&	2.24249	\\
1.7	&	0.066	&	0.1342	&	3.69515	&	3.1318	&	2.98999	\\
2.0	&	0.043	&	0.11535	&	4.34165	&	3.05612	&	4.58929	\\
\hline
\end{tabular}
\end{center}

\begin{center}
\tabcaption{Screening parameters of the $c\bar{c}$ (1p)state for $\mu$=$\mu_{c}$ for the different choices of $\nu$}\label{tt3}
\begin{tabular}{|cccccc|}
\hline
$\nu$	&	$\sigma$	&	$\mu_{c}$	&	$r_{0}$	&	M	&	$r_{D}$	\\
\hline
\hline
	&	$GeV^{\nu +1}$	&	GeV	&	fm	&	GeV	&	fm	\\
\hline											
0.1	&	0.728	&	0.92101	&	0.90081	&	3.43043	&	0.21426	\\
0.3	&	0.517	&	0.76931	&	1.03515	&	3.31203	&	0.25651	\\
0.5	&	0.385	&	0.63324	&	1.18696	&	3.24798	&	0.31163	\\
0.7	&	0.293	&	0.50668	&	1.35569	&	3.21827	&	0.38947	\\
0.9	&	0.227	&	0.42629	&	1.40402	&	3.32733	&	0.46292	\\
1.0	&	0.192	&	0.34172	&	1.58942	&	3.20186	&	0.57748	\\
1.1	&	0.177	&	0.30429	&	1.62205	&	3.22168	&	0.64852 \\
1.3	&	0.139	&	0.25813	&	1.61366	&	3.3257	&	0.76449	\\
1.5	&	0.11	&	0.18859	&	1.92708	&	3.22327	&	1.04639	\\
1.7	&	0.087	&	0.15148	&	2.24087	&	3.21433	&	1.30274	\\
2.0	&	0.062	&	0.11197	&	3.02153	&	3.19371	&	1.76243	\\
\hline
\end{tabular}
\end{center}

\end{multicols}
\includegraphics[width=6cm, height=6cm]{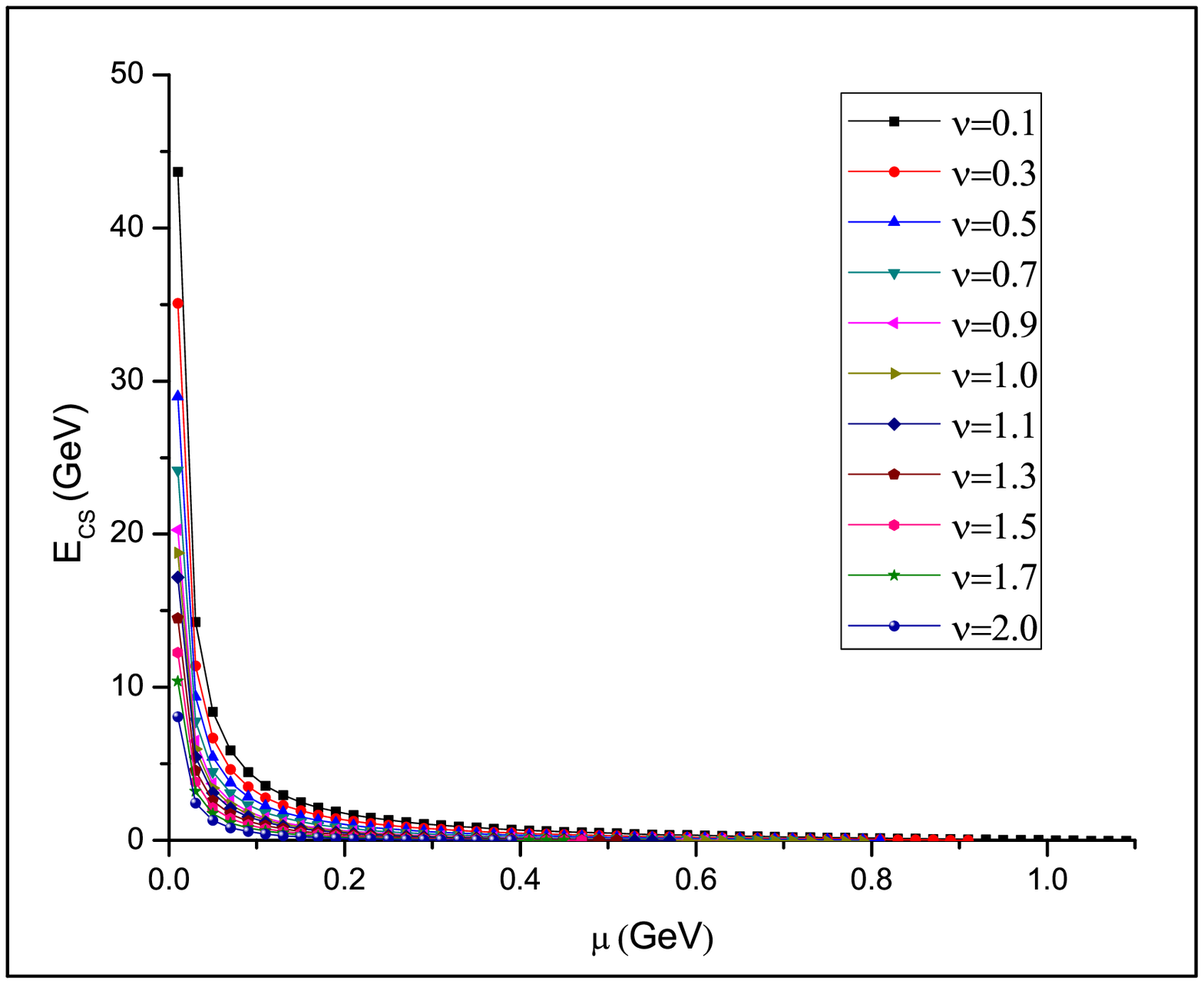}
\includegraphics[width=6cm, height=6cm]{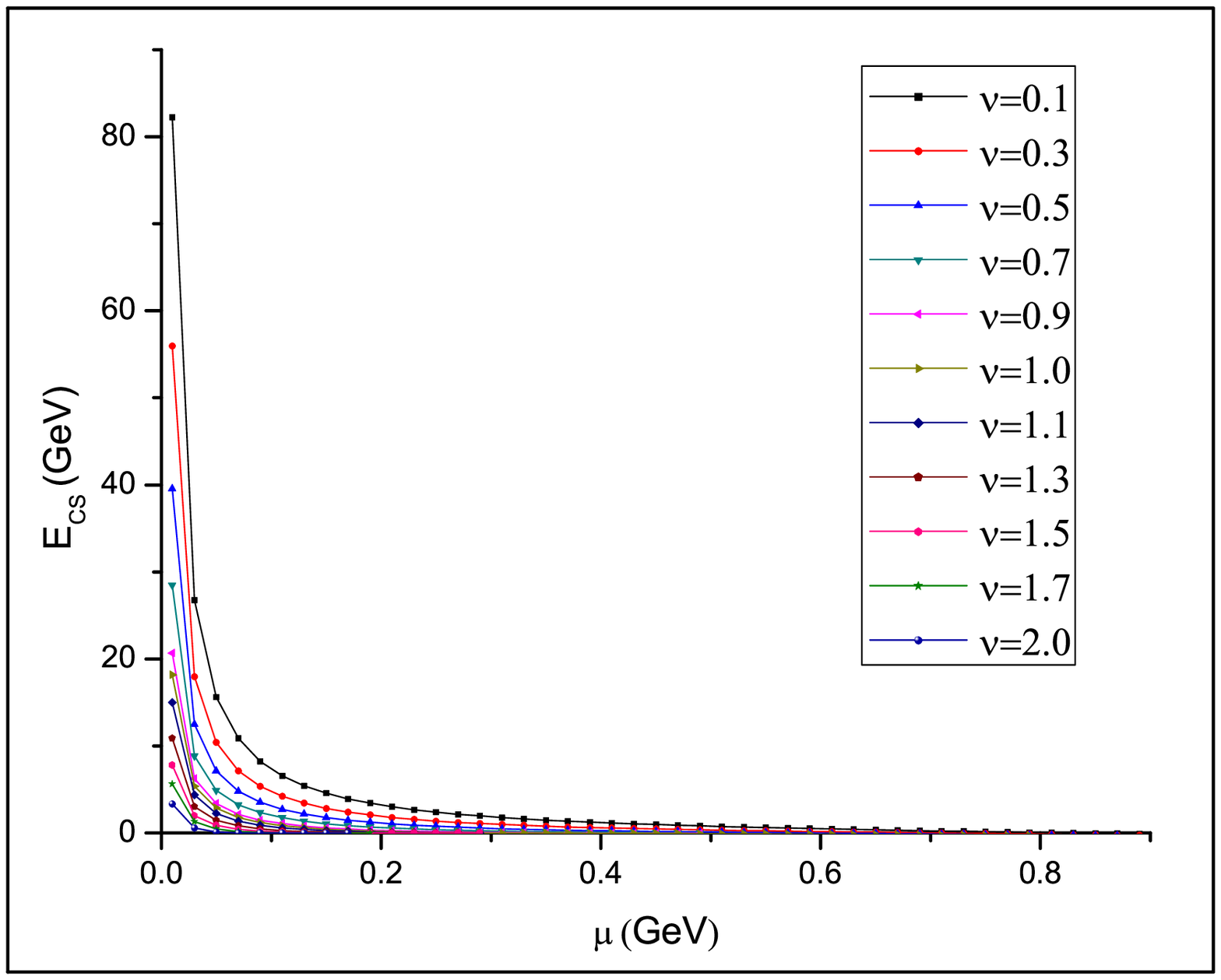}
\includegraphics[width=6cm, height=6cm]{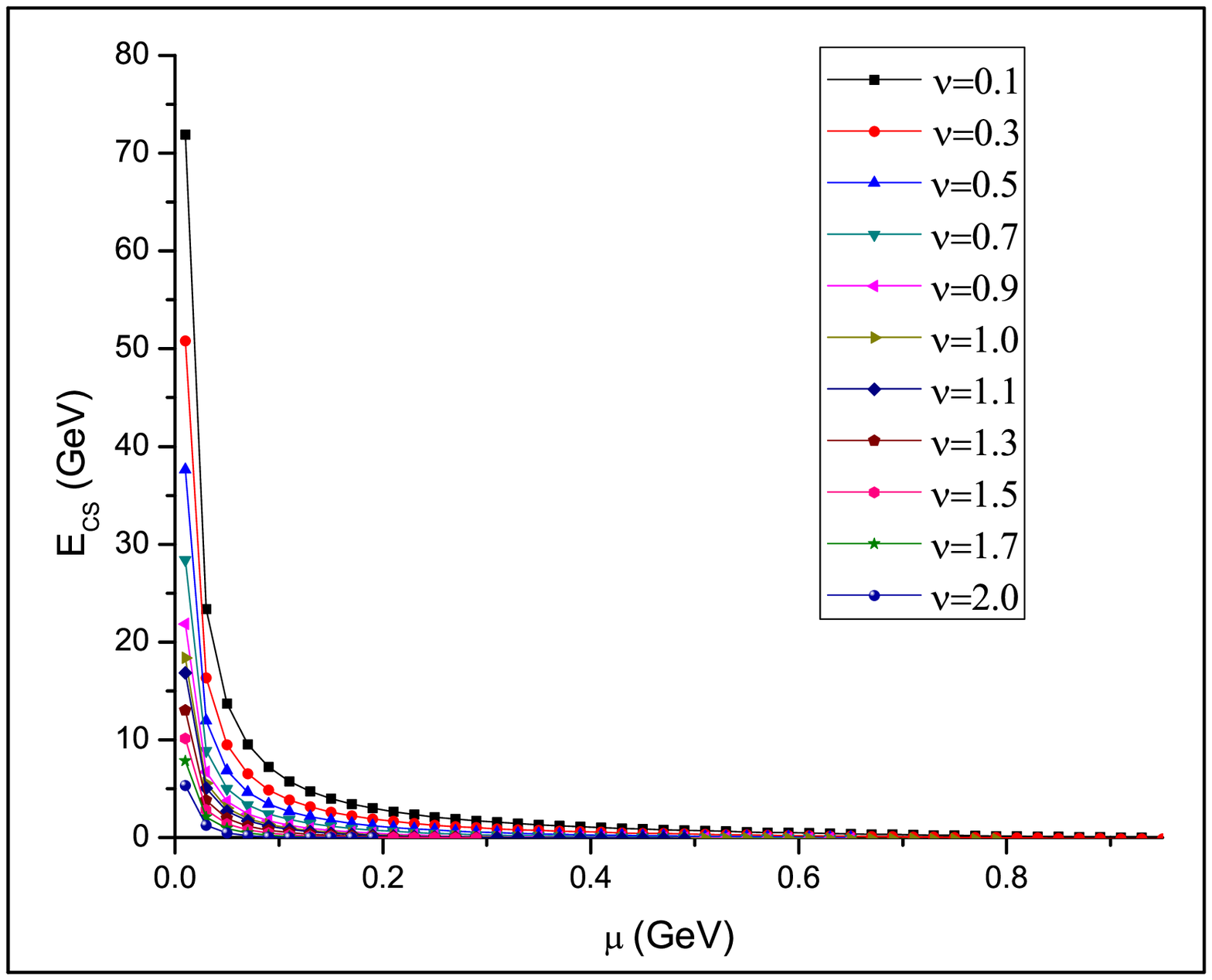}
\figcaption{Colour Screening energy $E^{n,l}_{cs}(\mu)$of the bound states of charmonium for the different values of $\nu$}\label{sigma_cc}

\begin{multicols}{2}

\begin{center}
\tabcaption{Screening parameters of the $b\bar{b}$ (1s)state for $\mu$=$\mu_{c}$ for the different choices of $\nu$}\label{tt4}
\begin{tabular}{|cccccc|}
\hline
$\nu$	&	$\sigma$	&	$\mu_{c}$	&	$r_{0}$	&	M	&	$r_{D}$	\\
\hline
\hline
	&	$GeV^{\nu +1}$	&	GeV	&	fm	&	GeV	&	fm	\\
\hline											
0.1	&	0.212	&	1.364	&	0.39804	&	9.64742	&	0.14467	\\
0.3	&	0.207	&	1.3374	&	0.39389	&	9.64676	&	0.14755	\\
0.5	&	0.203	&	1.31557	&	0.3896	&	9.6463	&	0.15000	\\
0.7	&	0.198	&	1.29695	&	0.38611	&	9.64467	&	0.15215	\\
0.9	&	0.193	&	1.2823	&	0.38369	&	9.6425	&	0.15389	\\
1.0	&	0.192	&	1.27767	&	0.38253	&	9.64227	&	0.15445	\\
1.1	&	0.188	&	1.2712	&	0.38228	&	9.63989	&	0.15523	\\
1.3	&	0.1821	&	1.26195	&	0.38184	&	9.6363	&	0.15637	\\
1.5	&	0.177	&	1.25525	&	0.3818	&	9.633	&	0.15721	\\
1.7	&	0.171	&	1.249	&	0.38231	&	9.6289	&	0.15799	\\
2.0	&	0.162	&	1.2408	&	0.38343	&	9.62256	&	0.15904	\\
\hline
\end{tabular}
\end{center}

\begin{center}
\tabcaption{Screening parameters of the $b\bar{b}$ (2s)state for $\mu$=$\mu_{c}$ for the different choices of $\nu$}\label{tt5}
\begin{tabular}{|cccccc|}
\hline
$\nu$	&	$\sigma$	&	$\mu_{c}$	&	$r_{0}$	&	M	&	$r_{D}$	\\
\hline
\hline
	&	$GeV^{\nu +1}$	&	GeV	&	fm	&	GeV	&	fm	\\
\hline											
0.1	&	0.511	&	1.06003	&	0.71679	&	9.97405	&	0.18616	\\
0.3	&	0.401	&	0.94723	&	0.79303	&	9.91534	&	0.20833	\\
0.5	&	0.323	&	0.83591	&	0.86467	&	9.8784	&	0.23607	\\
0.7	&	0.263	&	0.72545	&	0.93271	&	9.5453	&	0.27202	\\
0.9	&	0.217	&	0.6228	&	0.99571	&	9.84042	&	0.31685	\\
1.0	&	0.192	&	0.57257	&	1.03308	&	9.82733 &	0.34465	\\
1.1	&	0.179	&	0.534	&	1.06148	&	9.8272	&	0.36955	\\
1.3	&	0.149	&	0.4646	&	1.14111	&	9.8127	&	0.42475	\\
1.5	&	0.124	&	0.4128	&	1.23972	&	9.79239	&	0.47805	\\
1.7	&	0.103	&	0.3761	&	1.34601	&	9.76586	&	0.52470	\\
2.0	&	0.0784	&	0.3436	&	1.47302	&	9.72016	&	0.57433	\\
\hline
\end{tabular}
\end{center}

\begin{center}
\tabcaption{Screening parameters of the $b\bar{b}$ (1p)state for $\mu$=$\mu_{c}$ for the different choices of $\nu$}\label{tt6}
\begin{tabular}{|cccccc|}
\hline
$\nu$	&	$\sigma$	&	$\mu_{c}$	&	$r_{0}$	&	M	&	$r_{D}$	\\
\hline
\hline
	&	$GeV^{\nu +1}$	&	GeV	&	fm	&	GeV	&	fm	\\
\hline											
0.1	&	0.414	&	1.07177	&	0.69182	&	9.87827	&	0.18412	\\
0.3	&	0.336	&	0.94305	&	0.75084	&	9.84829	&	0.20925	\\
0.5	&	0.282	&	0.8238	&	0.80592	&	9.83431	&	0.23954	\\
0.7	&	0.239	&	0.7115	&	0.85072	&	9.8279	&	0.27735	\\
0.9	&	0.206	&	0.61433	&	0.87292	&	9.82732	&	0.32122	\\
1.0	&	0.192	&	0.5725	&	0.87716 &	9.82728	&   0.34469	\\
1.1	&	0.178	&	0.5349	&	0.88272	&	9.82477	&	0.36892	\\
1.3	&	0.155	&	0.4731	&	0.90015	&	9.81962	&	0.41712	\\
1.5	&	0.136	&	0.4249	&	0.93709	&	9.81207	&	0.46443	\\
1.7	&	0.119	&	0.3857	&	0.99319	&	9.80052	&	0.51163	\\
2.0	&	0.098	&	0.34075	&	1.1046	&	9.77959	&	0.57913	\\
\hline
\end{tabular}
\end{center}

Table (\ref{tt1}) to (\ref{tt6}) shows the screening parameter of 1s, 2s, and 1p-state of the $c\bar{c}$ and $b\bar{b}$ and fig (\ref{sigma_cc}) and (\ref{sigma_bb}) shows the change in energy eigenvalue with respect to screening parameter $\mu$ for the different choices of potential exponent $\nu$. The value of $\mu_{c}$ is extracted from the condition given by eqn(\ref{critical_mu}). It is observed that the critical value for the screening mass $\mu_{c}$ decreases with increase in the choice of potential exponent $\nu$. Also with increase of $\nu$, the color screening radii $r_{D}$ ($r_{D}$=1/$\mu_{c}$)and the r.m.s value ($r_{0}$) at the last binding energy of the quarkonia state at $\mu$=$\mu_{c}$ show a increasing trend. The mass of the quarkonia state at $\mu$=$\mu_{c}$ is tabulated below the column represented by M. The resultant colour screening parameter, $\mu_{c}$ for each of the quarkonia states against the potential exponent, $\nu$ are plotted in fig. (\ref{mu_nu1}).
\end{multicols}
\includegraphics[width=6cm, height=6cm]{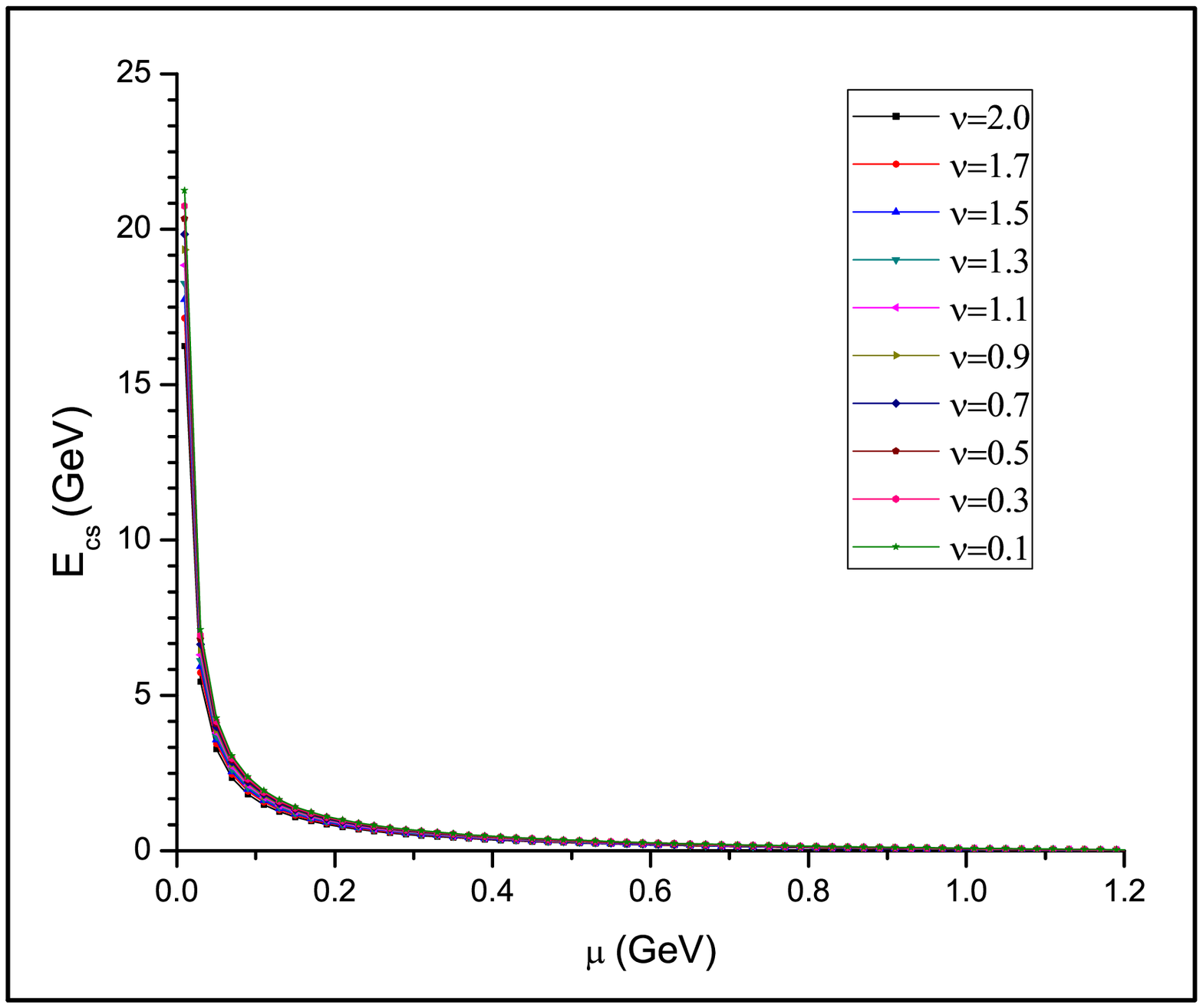}
\includegraphics[width=6cm, height=6cm]{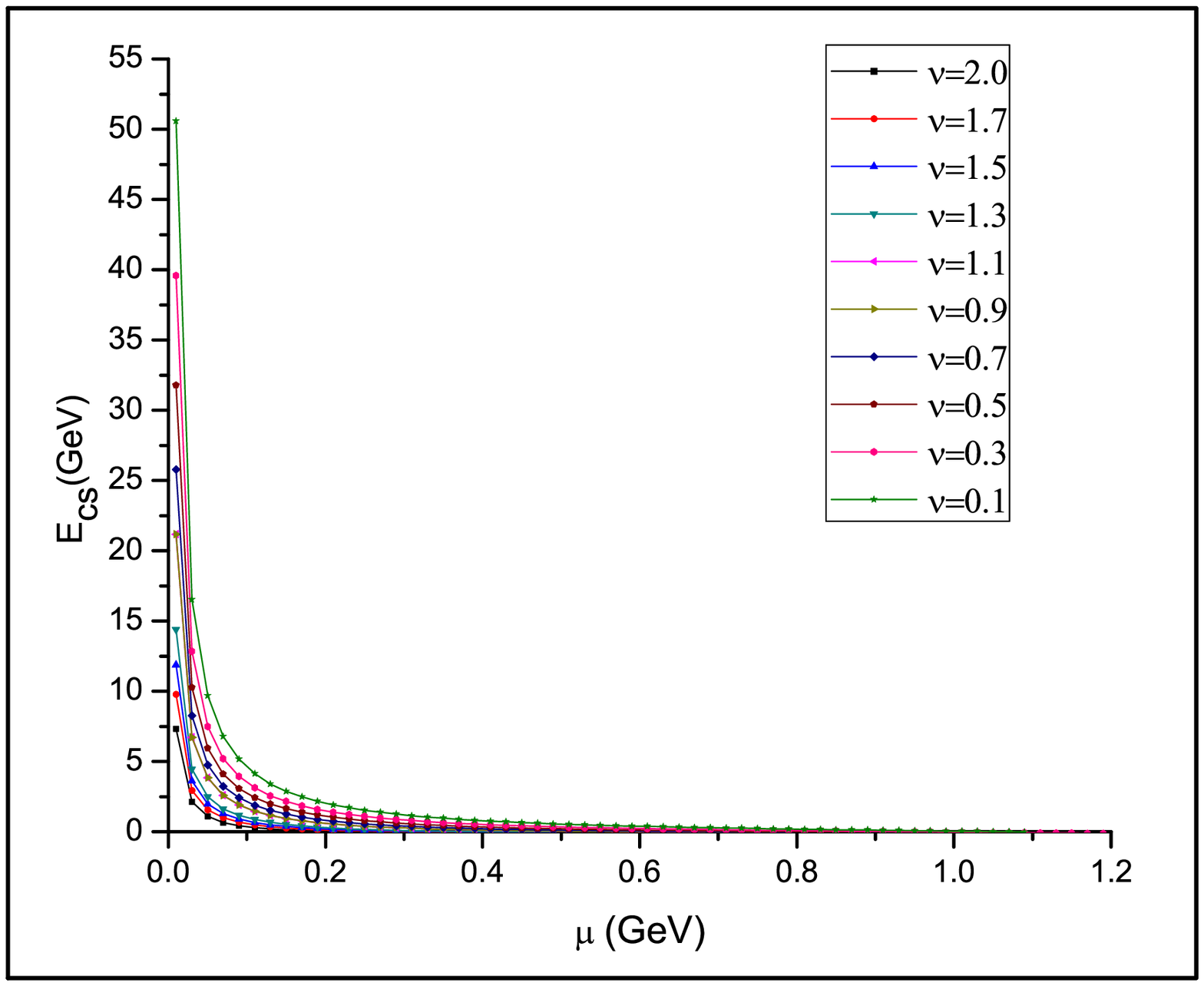}
\includegraphics[width=6cm, height=6cm]{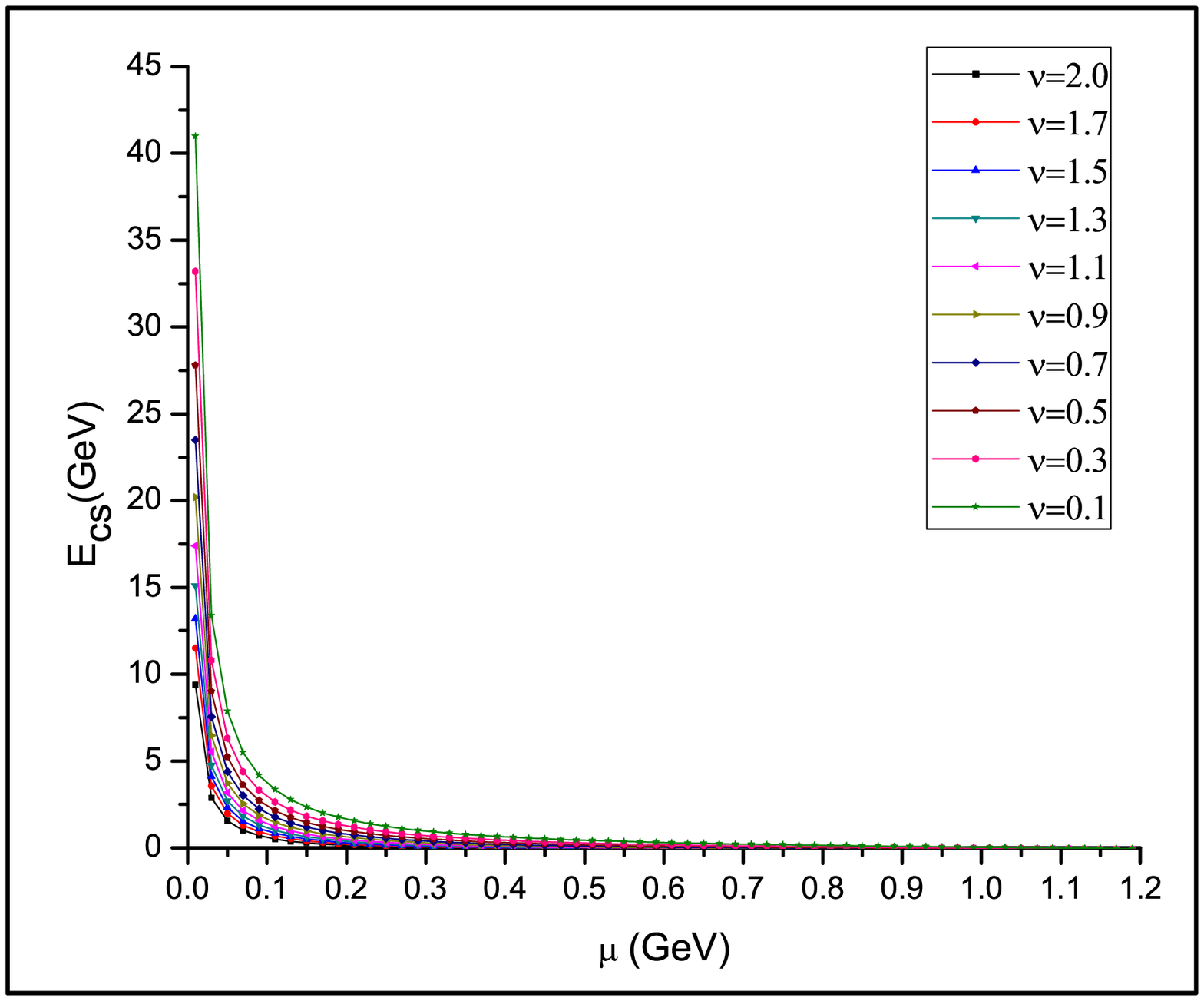}
\figcaption{Colour Screening energy $E^{n,l}_{cs}(\mu)$ of the bound states of bottomonium for the different values of $\nu$}\label{sigma_bb}

\begin{multicols}{2}

\begin{center}
\includegraphics[width=6cm, height=6cm]{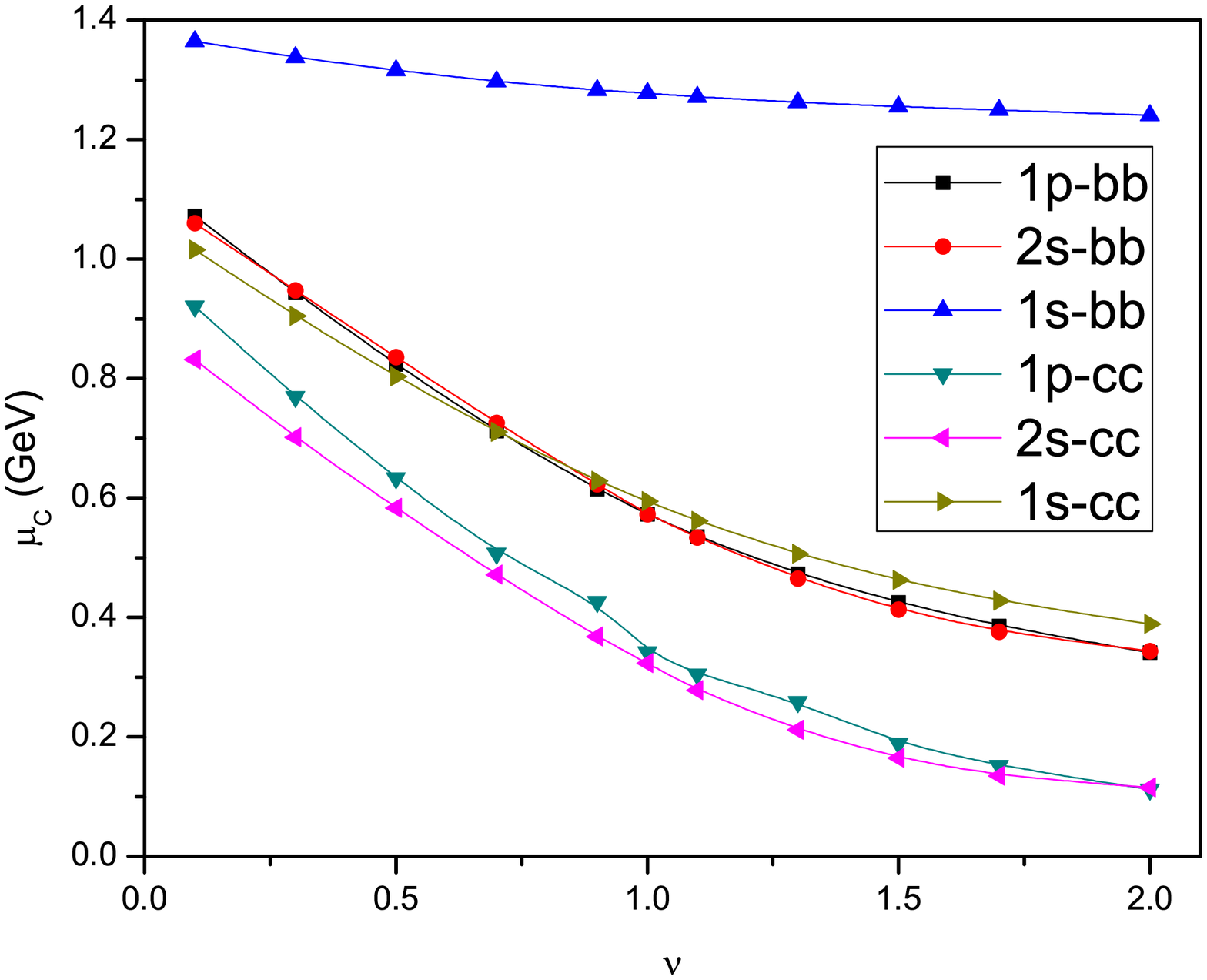}
\figcaption{Colour screening parameter $\mu_{c}$ with increase in power exponent $\nu$}\label{mu_nu1}
\end{center}

\section{The Vacuum Screening Mass}

At T=0, The absence of light quarks indicates the screening parameter $\mu$=0 while the presence of light quark-antiquark from vacuum correspond to $\mu$(T=0)$\neq$0. As the separation between Q-$\bar{Q}$ increases the gluonic flux that binds Q and $\bar{Q}$ breaks and the light quark and antiquark pairs are produced out of vacuum.This breaking of string is attributed to the creation of $q\bar{Q}$ and $\bar{q}Q$ but not exactly due to colour screening. Energy is required to bring out the virtual $q\bar{q}$ pair from vacuum and hence, $\mu(T=0)\neq$0.

Considering the vacuum screening, the effective binding energy can be represented as \cite{Karsch}
\begin{equation}\label{e_vs}
E_{vs}(T=0)=2m_{Q\bar{q}}-M_{Q\bar{Q}}
\end{equation}

where, $m_{Q\bar{q}}$ is mass of heavy-light quark and $m_{Q\bar{Q}}$ is mass of state of $c\bar{c}$ and ${b\bar{b}}$. Here, we consider $m_{Q\bar{q}}$ as $D_{0}$ for charmonia and $B_{0}$ for bottomonia.



comparing eqn. (\ref{e_vs}) with eqn. (\ref{e_cs}), the vacuum screening parameter, $\mu_{vs}$ has been calculated for the different choices of power exponent $\nu$ and results are tabulated in table (\ref{charm11}) and (\ref{bottom11}) for charmonia and bottomonia states respectively. The vacuum screening parameter, $\mu_{vs}$ obtained for each of the quarkonia states are potted against the potential exponent, $\nu$ in fig(\ref{mu_nu2}).

\begin{center}
\tabcaption{Vacuum screening parameter $\mu_{vs}$ for charmonia states}\label{charm11}
\begin{tabular}{|cccc|}
\hline
   &	$J/\Psi$	&	$\Psi^{'}$	&	$\chi_{c}$	\\
\hline
\hline
$\nu$	&		&	$\mu_{vs}$(GeV)	&		\\	
\hline
\hline
0.1	&	0.4009	&	0.7659	&	0.6009	\\
0.3	&	0.3229	&	0.5243	&	0.4269	\\
0.5	&	0.2673	&	0.3737	&	0.3178	\\
0.7	&	0.2237	&	0.2714	&	0.2420	\\
0.9	&	0.1884	&	0.1994	&	0.1873	\\
1.0	&	0.1743	&	0.1759	&	0.1621	\\
1.1	&	0.1599	&	0.1472	&	0.1461	\\
1.3	&	0.1356	&	0.1094	&	0.1148	\\
1.5	&	0.1155	&	0.0812	&	0.0908	\\
1.7	&	0.0982	&	0.0607	&	0.0719	\\
2.0	&	0.0772	&	0.0394	&	0.0511	\\
\hline
\end{tabular}
\end{center}

\begin{center}
\tabcaption{Vacuum screening parameter $\mu_{vs}$ for bottomonia states}\label{bottom11}
\begin{tabular}{|cccc|}
\hline
 & $\Upsilon$ & $\Upsilon^{'}$ & $\chi_{b}$	\\
\hline
\hline
$\nu$	  &		 &	$\mu_{vs}$(GeV)  &		\\
\hline
\hline
0.1	&	0.2019	&	0.3151	&	0.2756	\\
0.3	&	0.1973	&	0.2473	&	0.2237	\\
0.5	&	0.1933	&	0.1991	&	0.1876	\\
0.7	&	0.1886	&	0.1623	&	0.1591	\\
0.9	&	0.1838	&	0.1338	&	0.1371	\\
1.0	&	0.1826	&	0.1192	&	0.1277	\\
1.1	&	0.1790	&	0.1104	&	0.1185	\\
1.3	&	0.1735	&	0.0918	&	0.1032	\\
1.5	&	0.1686	&	0.0764	&	0.0905	\\
1.7	&	0.1629	&	0.0635	&	0.0792	\\
2.0	&	0.1543	&	0.0483	&	0.0652	\\
\hline
\end{tabular}
\end{center}

\begin{center}
\includegraphics[width=6cm, height=6cm]{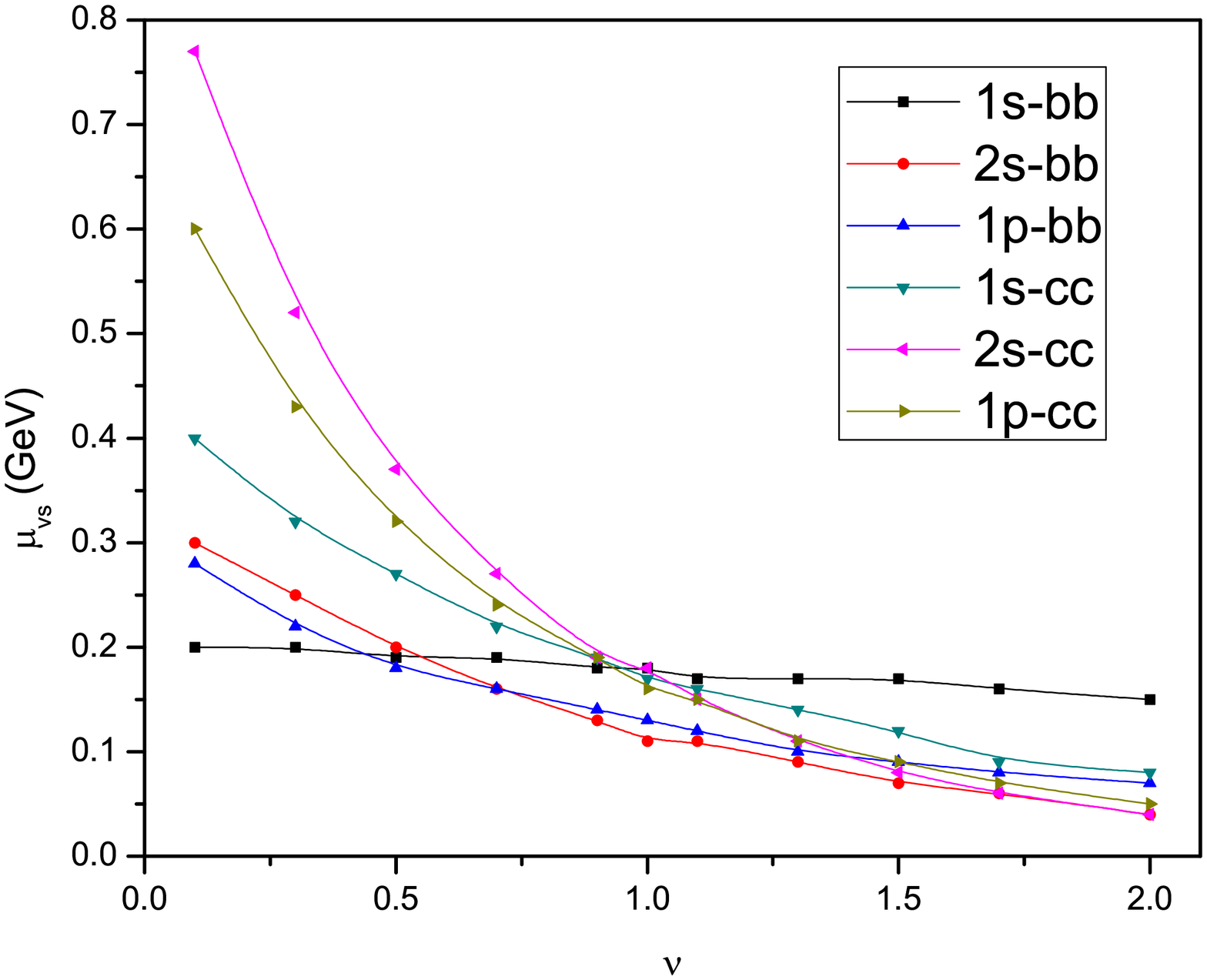}
\figcaption{Vacuum screening parameter $\mu_{vs}$ with increase in power exponent $\nu$}\label{mu_nu2}
\end{center}

\section{Conclusion}

The effect of medium on the binding energy of the $Q\bar{Q}$ states are studied by introducing a medium dependent screening mass parameter, $\mu$. For the different choices of $\mu$ we have calculated the effective binding energy and the bound state radii by solving the Schr$\ddot{o}$dinger equation. The effective binding energy ($E_{cs}^{n,l}$) as defined in eqn.(\ref{e_cs}) is found to vanish at a particular value of $\mu=\mu_{c}$. This value $\mu_{c}$ is then defined as critical screening mass parameter of the quarkonia state. Above this value bound state will not be possible. Corresponding to this critical screening mass parameter,$\mu_{c}$ we obtained the screening radii $r_{D}= 1/\mu_{c}$ The corresponding r.m.s. radius of the quarkonium state $r_{0}$at $\mu$=$\mu_{c}$ is also computed for each choices of $\nu$. A part from the medium screening effects, the vacuum screening effects by considering $\mu$(T=0)$\neq$0 is also studied according to eqn.(\ref{e_vs}). Here like $\mu_{c}$ we obtain the vacuum screening parameter, $\mu_{vs}$ for each choices of $\nu$.
\end{multicols}
\begin{center}
\includegraphics[width=5cm, height=6cm]{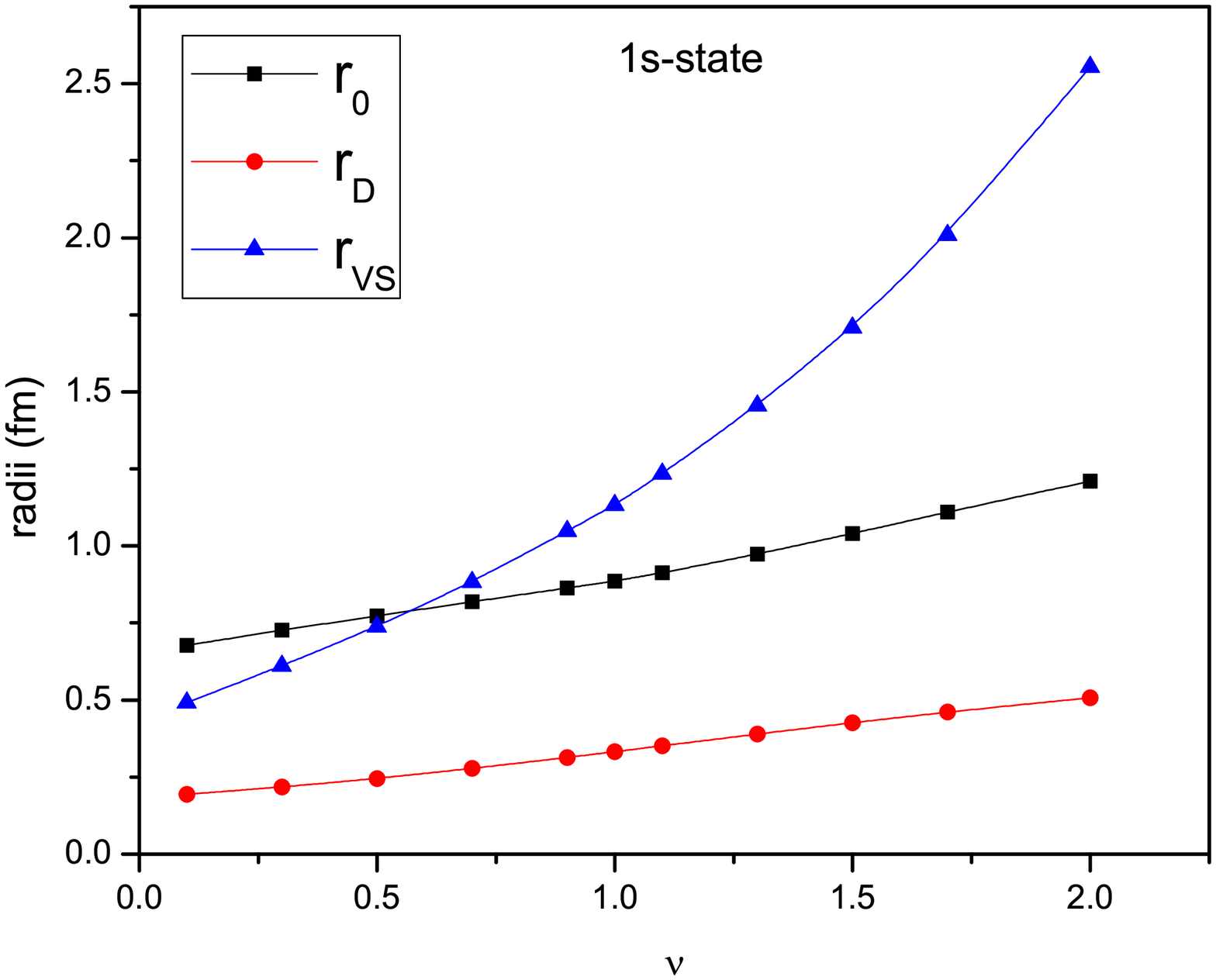}
\includegraphics[width=5cm, height=6cm]{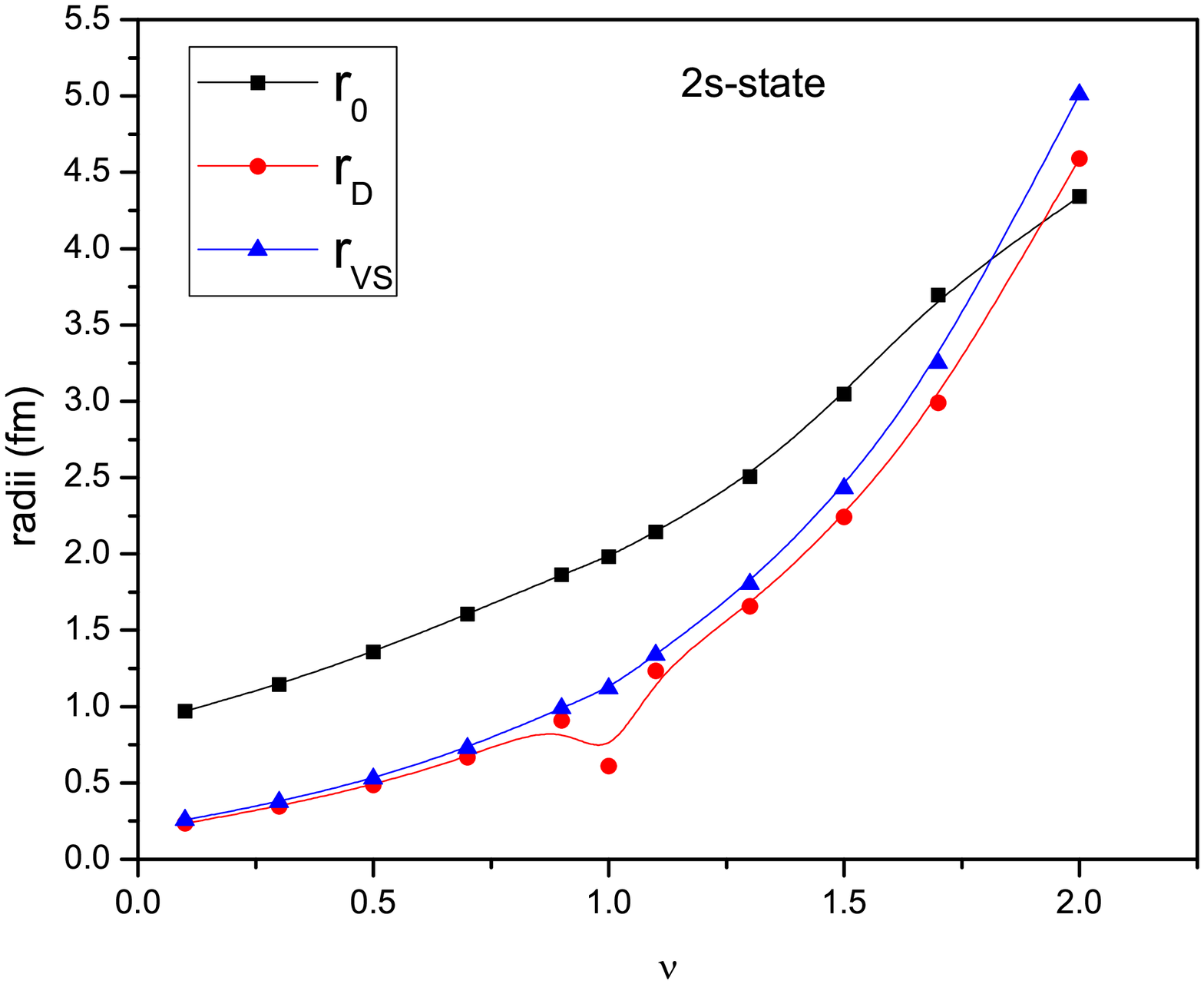}
\includegraphics[width=5cm, height=6cm]{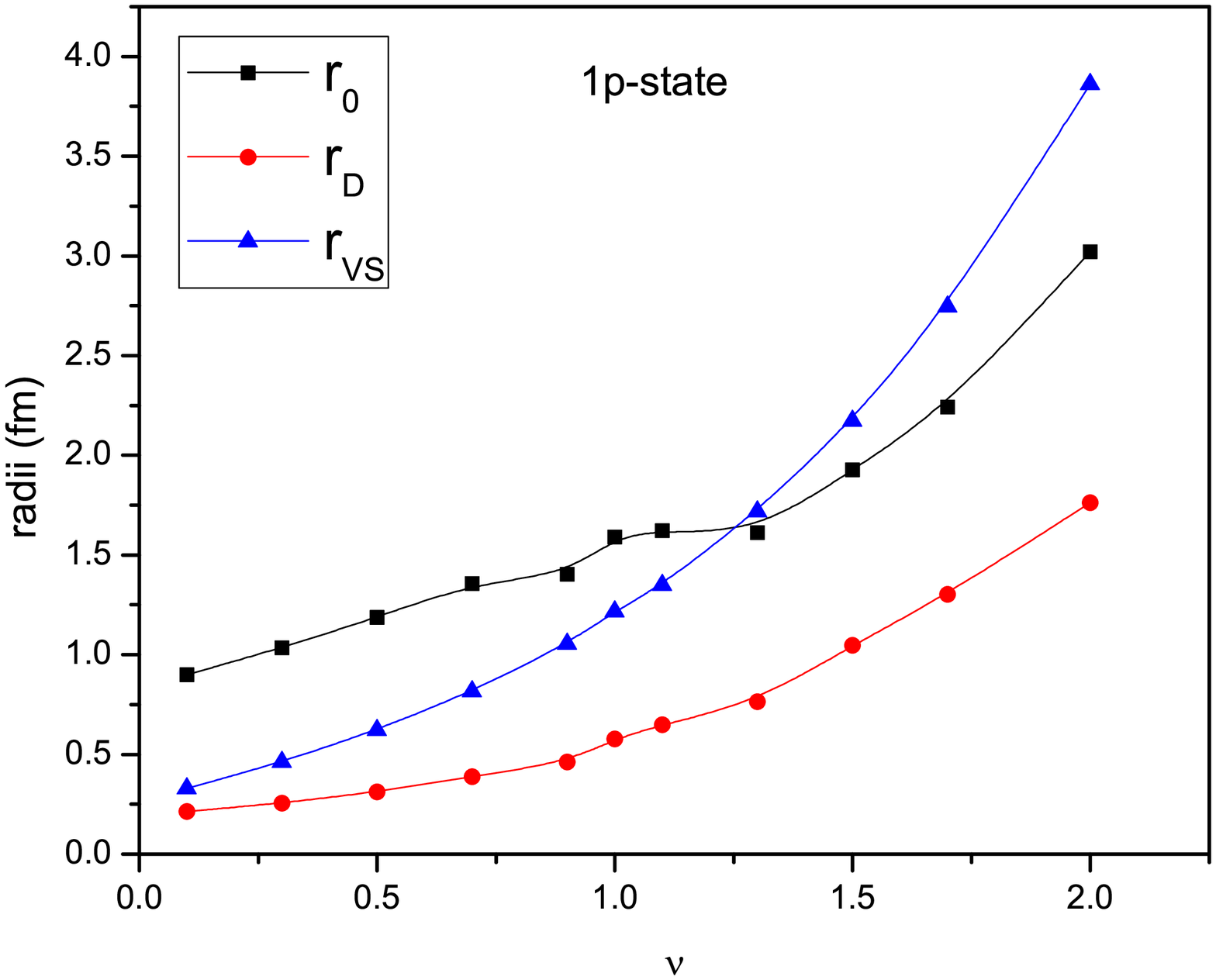}
\figcaption{Radii with increase in power exponent $\nu$ for charmonia for 1s, 2s and 1p-states respectively}\label{radi_nu1}
\end{center}

\begin{center}
\includegraphics[width=5cm, height=6cm]{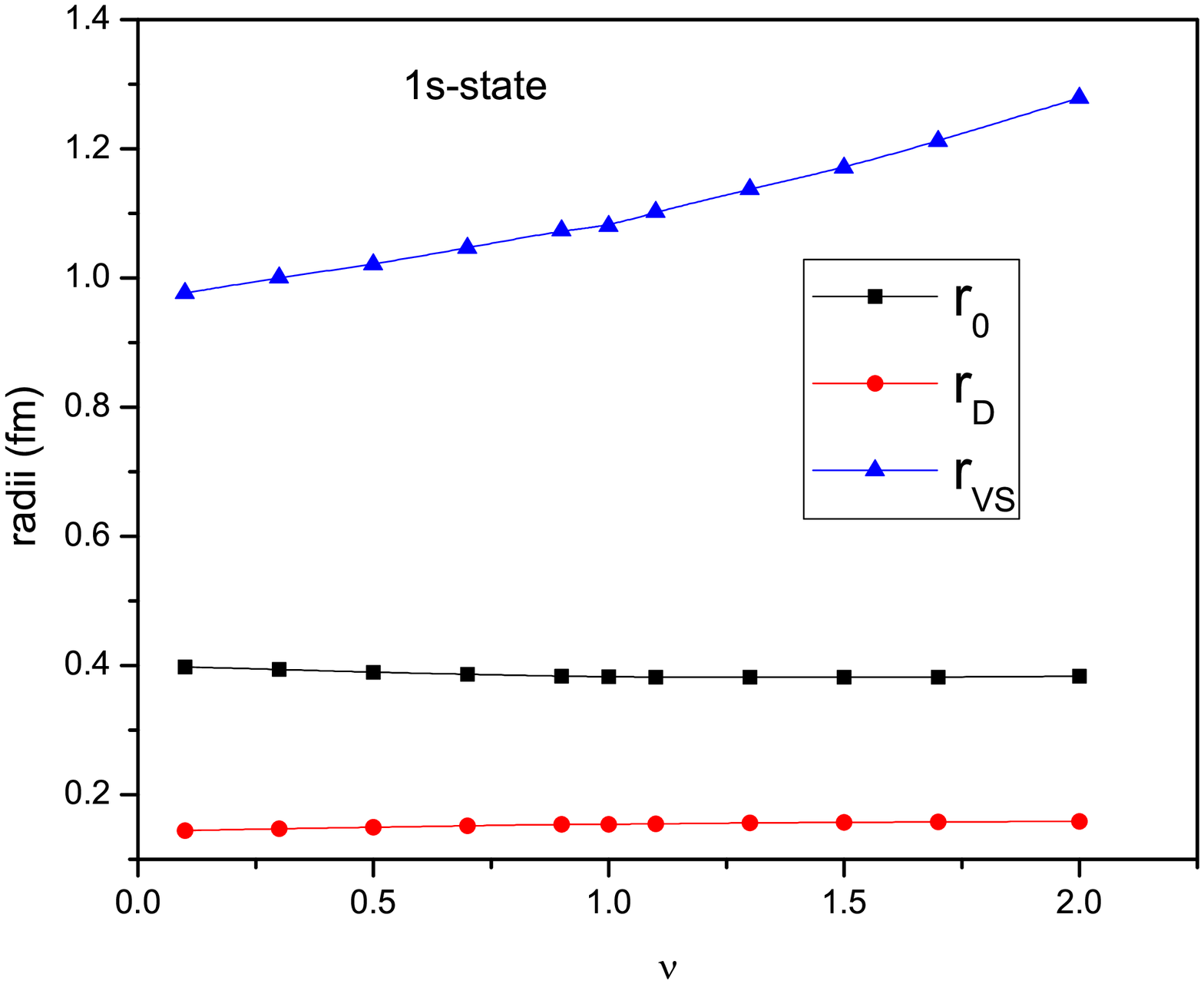}
\includegraphics[width=5cm, height=6cm]{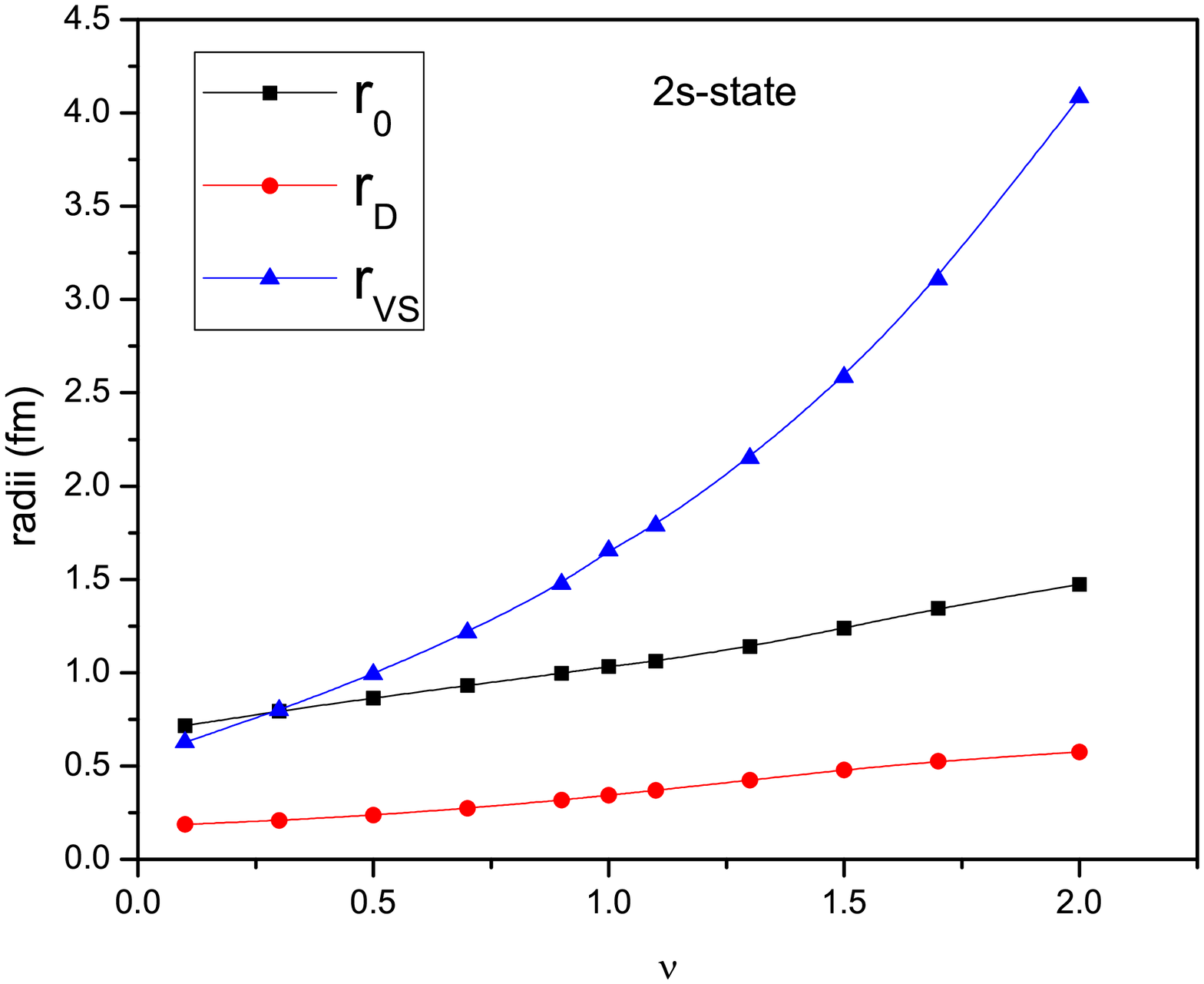}
\includegraphics[width=5cm, height=6cm]{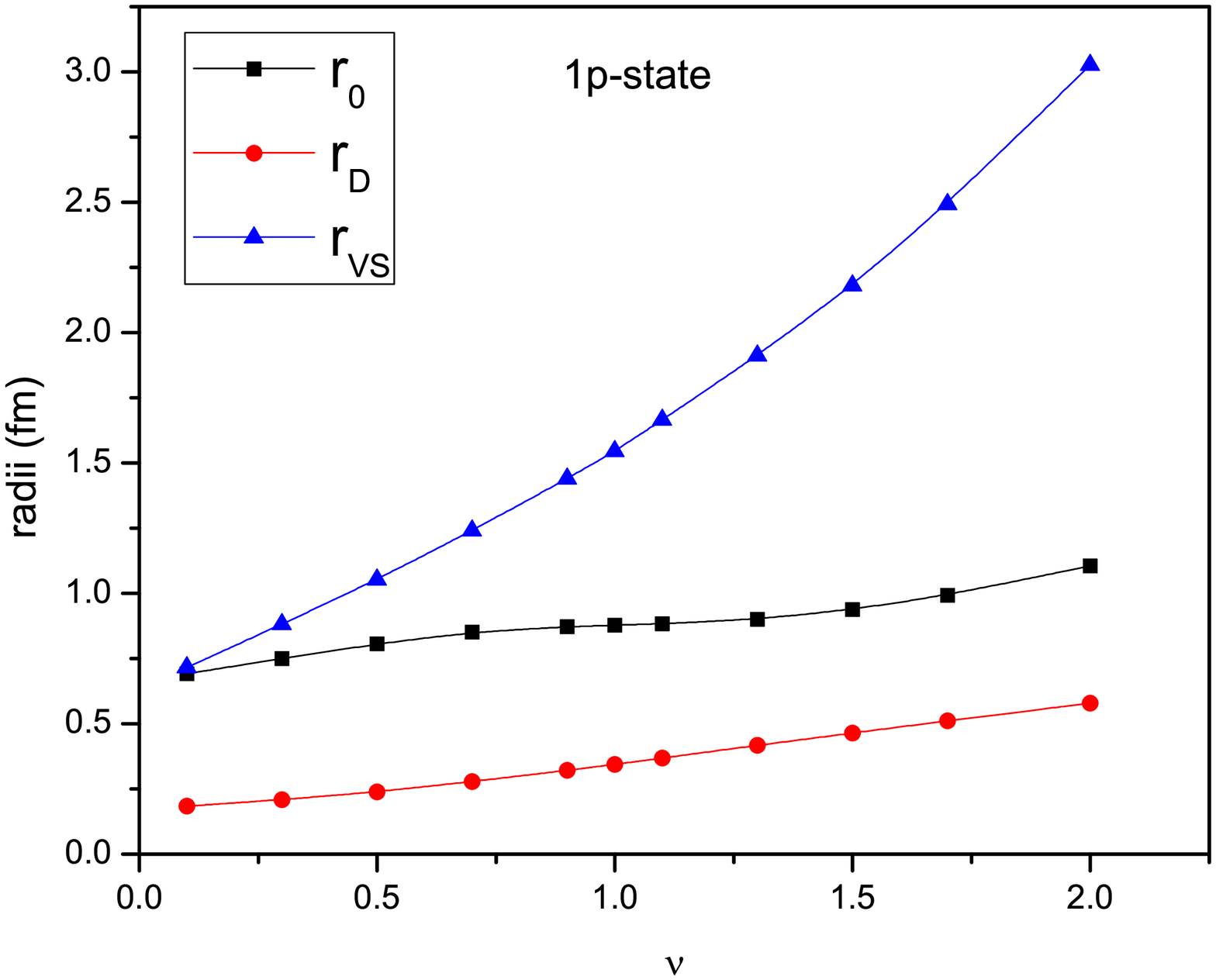}
\figcaption{Radii with increase in power exponent $\nu$ for bottomonia for 1s, 2s and 1p-states respectively}\label{radi_nu2}
\end{center}

\begin{multicols}{2}

\begin{center}
\includegraphics[width=6cm, height=6cm]{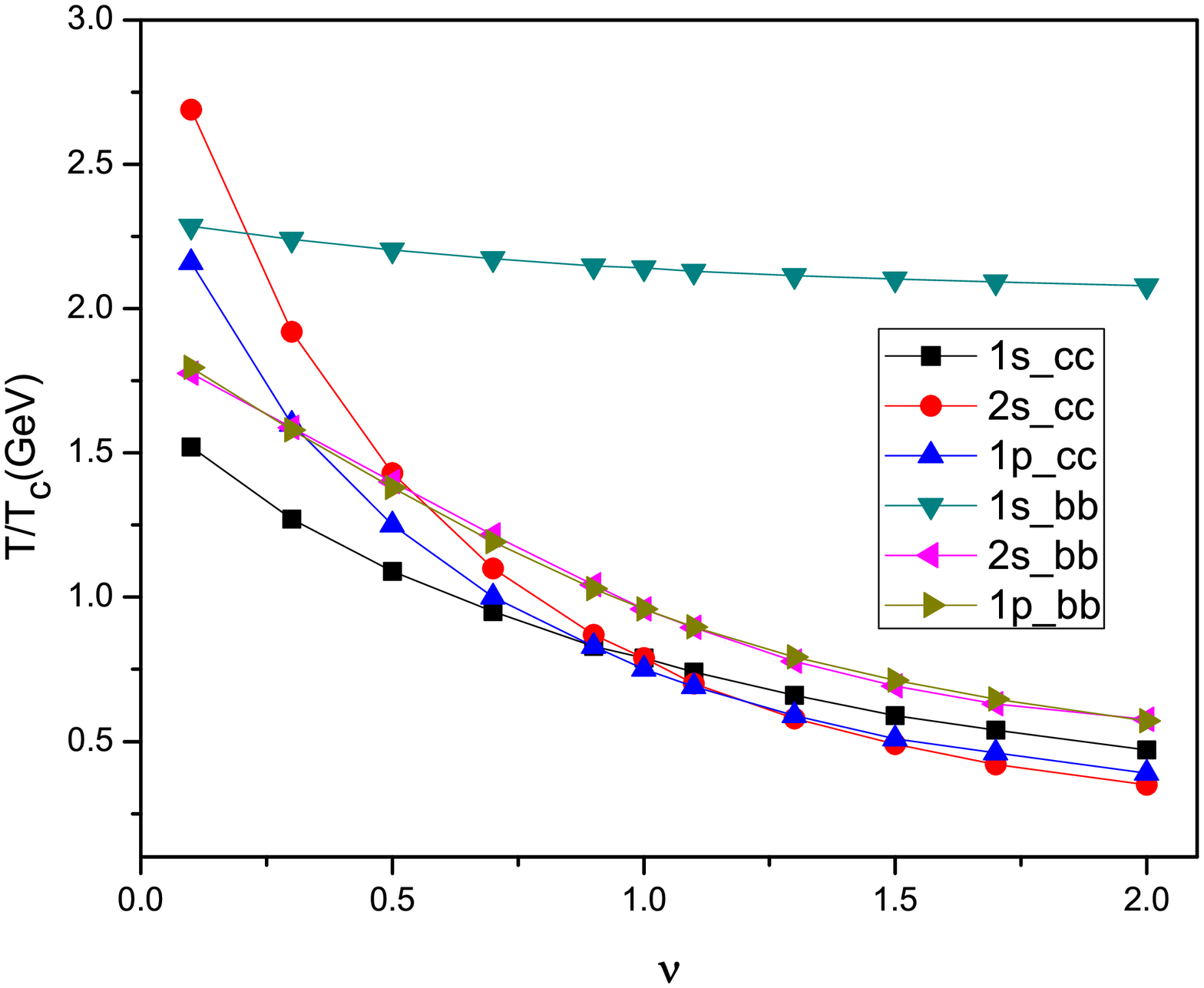}
\figcaption{Temperature at critical colour screening parameter ($\mu_{c}$) with increase in power exponent $\nu$}\label{T_nu1}
\end{center}

From fig.(\ref{mu_nu1}) and (\ref{mu_nu2}) one can conclude that $\mu_{c}$ and $\mu_{vs}$ decreases with increase in the potential exponent $\nu$ in all the cases while for a particular choices of $\nu$ it increases for excited states. And the vacuum screening radii ($r_{vs}$=1/$\mu_{vs}$)increases with increase in potential exponent $\nu$.

The vacuum screening of the 1s-state of charmonia will be stable above $\nu$=0.5 similarly for 2s-state and 1p-state of charmonia vacuum screening will be stable above $\nu$=1.7 and 1.3 respectively. Similarly for the case of bottomonia 1s-state, 2s-state and 1p-state will be stable against vacuum screening. While these states will be unstable due to colour screening in both charmonia and bottomonia cases studied here.


Fig.(\ref{radi_nu1}) and (\ref{radi_nu2}) shows a plot of radii (i.e last banding radii $r_{0}$, colour screening radii $r_{D}$ and vacuum screening radii $r_{VS}$) of different (i.e. 1s, 2s and 1p) states of charmonia and bottomonia with different choices of potential exponent $\nu$. The different states of charmonia and bottomonia get screened, which represents their instability in medium.

\begin{center}
\tabcaption{Colour screening temperature for charmonia}\label{dc1}
\begin{tabular}{|cccc|}
\hline
$\nu$	&	&	$T/T_{c}$	&	\\
\hline
\hline
	&	1s-state &	2s-state &	1p-state\\
\hline
0.1	&	3.50	&	2.90	&	3.19	\\
0.3	&	3.14	&	2.49	&	2.71	\\
0.5	&	2.82	&	2.11	&	2.26	\\
0.7	&	2.51	&	1.75	&	1.86	\\
0.9	&	2.25	&	1.41	&	1.60	\\
1.0	&	2.14	&	1.27	&	1.33	\\
1.1	&	2.03	&	1.12	&	1.21	\\
1.3	&	1.86	&	0.91	&	1.06	\\
1.5	&	1.71	&	0.76	&	0.83	\\
1.7	&	1.61	&	0.66	&	0.71	\\
2.0	&	1.48	&	0.59	&	0.59	\\
\hline											
\end{tabular}
\end{center}

\begin{center}
\tabcaption{Colour screening temperature for bottomonia}\label{db1}
\begin{tabular}{|cccc|}
\hline
$\nu$	&	&	$T/T_{c}$	&	\\
\hline
\hline
	&	1s-state &	2s-state &	1p-state\\
\hline
0.1	&	4.62	&	3.64	&	3.68	\\
0.3	&	4.54	&	3.28	&	3.26	\\
0.5	&	4.45	&	2.92	&	2.88	\\
0.7	&	4.41	&	2.56	&	2.52	\\
0.9	&	4.36	&	2.23	&	2.21	\\
1.0	&	4.35	&	2.07	&	2.07	\\
1.1	&	4.32	&	1.95	&	1.95	\\
1.3	&	4.29	&	1.72	&	1.75	\\
1.5	&	4.27	&	1.55	&	1.59	\\
1.7	&	4.25	&	1.44	&	1.47	\\
2.0	&	4.23	&	1.33	&	1.32	\\
\hline											
\end{tabular}
\end{center}

Table (\ref{dc1}) and (\ref{db1}) represents the temperature at the critical colour screening parameter $\mu_{c}$ which is shown in fig.(\ref{T_nu1}) for the different charmonia and bottomonia states for the different choices of $\nu$, which is calculated using screening parameter $\mu_{c}$ \cite{Agnes},
\begin{equation}
\mu_{c}(T)=0.24 + 0.31(\frac{T}{T_{0}}-1)
\end{equation}

The survival probability of the J/$\psi$, $\psi'$ and $\chi_{c}$-state of charmonia has been obtained as 2-2.5$T_{c}$, 1.1$T_{c}$ and 1.1-1.3$T_{c}$ \cite{Shuryak,Alberico} while for $\Upsilon$ $\sim$3-4$T_{c}$\cite{Digal,Mocsy}.Here our results agree for the charmonia case for the choice of exponent $\nu\sim$1.1. However, for the bottomonia case our results overestimate with the prediction of\cite{Digal,Mocsy} for all potential exponents studied here.

We conclude here, the in medium properties of charmonia states  can be studied by cornel like potential ($\nu\sim1.0$) or by slightely higher than it. While in case of bottomonia states more detailed study is needed.


\end{multicols}

\clearpage

\end{document}